# Using utility graphs to search for Pareto-optimal outcomes in complex, interdependent issue negotiations

Valentin Robu[1,2,3*], Mark Klein[3,4]
[1] CWI, The Netherlands National Research Center for Mathematics & Computer Science, Amsterdam
[2] Eindhoven University of Technology (TU/e), Eindhoven, The Netherlands
[3] Massachusetts Institute of Technology (MIT), Center for Collective Intelligence, Cambridge, MA, USA
[4] School of Collective Intelligence, UM6P (University Mohammed VI Polytechnic), Rabat Morocco
Emails: v.robu@cwi.nl, m_klein@mit.edu

**Abstract:** This paper studies how utility graphs decomposition algorithms can be used to effectively search for Pareto-efficient outcomes in complex automated negotiation. We propose a number of algorithms that can efficiently handle high-dimensional utility graphs, and test them on a variety of utility graph topologies, generated based on state of the art methods for analysing complex graphs. We show that we can achieve exponential speed-up, for many structures, even for very large utility graphs. To our knowledge, our approach can handle the largest utility spaces to date for complex interdependent negotiations, in terms of number of issues. Moreover, we examine the performance of our algorithms across two different types of elicitation queries from the literature: value and comparison queries, thus making a connection between automated negotiation and the preference elicitation literature.

## 1. Introduction

Negotiation is a powerful tool for modelling complex interactions between self-interested agents, which can be people, companies or increasingly, AI-enabled autonomous agents, that aim to reach the best agreement for their human owners. While negotiation is often thought as a competitive process, in which one party wins and the other one loses, in practice most real negotiations involve more complex, win-win scenarios (Raiffa [20]), in which agreements can be found that maximize the utilities of both agents. Such outcomes (agreements) are called Pareto-efficient, i.e. it is not possible to find another outcome that would increase one agent's utility, without making another agent worse off.

Yet, finding agreements that are Pareto-efficient is a challenging computational problem, especially in complex negotiation domains, where issues negotiated upon are interdependent (i.e. the utility of the value chosen for one negotiation issue depends strongly on the choice for other ones). Consider, for example, the negotiations between parties in a logistic supply chain: producers want to have certain combinations of resources/quantities, delivered at certain times to be able to produce their goods, while suppliers may face similar constraints in their cost function for supplying different combinations of items. Or the peer-to-peer negotiations between prosumers in a decentralised power grid, that require certain amounts of energy at different times and locations, which involve non-linear constraints, especially if the capacity of the distribution network is limited. Yet another example where automated agents are used are negotiations between advertisers and news websites, for determining which ads to display to each user. All these examples involve complex trade-offs between multiple issues, and determining the best outcome/deal involves solving a challenging computational problem, that takes into account the (potentially opposing) preferences of the parties. In the field of multi-agent systems and distributed AI, this has given rise to the field of complex automated negotiation, that aims to study how an efficient outcome can be found in these domains with limited computation capacity.

An important concept in all such negotiations is the utility function of the agents. Most utility models developed in the automated negotiation literature assume that utility functions are linearly additive, i.e.

* Research performed during the author's visiting appointment at the Center for Collective Intelligence at MIT.

the utility of the agents over multiple issues can simply be expressed as a weighted sum between the issues being negotiated on. However, this approach is often too simple to model the complexity of real negotiation domains. Hence, a lot of research interest in the distributed AI community focuses on complex negotiations, that consider *utility interdependencies* between different issues. A variety of approaches have been proposed for this problem. In this vein, Klein et al. [14] consider negotiation over a set of binary issues (similar to this paper), and look at the design of a mediator agent that uses simulated annealing to efficiently explore the negotiation space. The work of Fujita et al. [11], Ito et al. [12], as well as Marsa Maestre et al. [17] propose a preference representation in the form of multi-dimensional intervals, which can give rise to a highly non-linear utility space. They consider design of a mediator agent that can efficiently explore this space to propose mutually profitable deals for both parties, using learning techniques such as simulated annealing, tabu search or evolutionary computation. This representation is also being explored as part of the international negotiating agent competition (Aydogan et al. [1]). Klein proposes building a utility function compiler for this problem, using concepts from computational geometry [15], while other works have looked at evolutionary computation as a means of determining Pareto-optimal outcomes in a complex utility space [7]. By contrast, in this work, we aim to leverage the power of graphical representations (specifically utility graphs) to model complex negotiations. Which such representations have been considered in prior work of Robu et al. [21] and Fujita et al. [22], to our knowledge no paper has proposed a general-purpose graph decomposition algorithm to address very large-scale utility domains. In this paper, we aim to build a unified formalism and graphical representation for this problem, and show how it can be used to handle negotiations with a truly large number of issues - much larger than has been possible in Pareto search in previous approaches to automated negotiation with interdependent-value issues.

Graphs (in our case hypergraphs, allowing multi-vertex edges) are natural model of preferences in many domains where automated negotiation is used. Complex systems science discusses a number of ways to generate graph structures, that yield graphs with very different connectivity and underlying structure, such as random, small-world and scale-free graphs. Yet, these insights have never been explored in automated negotiation, because efficient utility representations and algorithms to model large-scale preference graphs were not developed before. Our work aims to fill this knowledge gap, opening a new avenue for handling complex-interdependent negotiations.
Moreover, our work also makes a key link between negotiation and preference elicitation of multi-issue preferences, a link that has not been made before, in a formal interdependent-value issues model. In particular, by considering mediated negotiation, we can explore how different types of queries can be used by the mediator to find Pareto-efficient agreement between negotiating parties, as well as the costs and trade-offs between different preference types.

## 2. Models of interdependent utility functions

In order to represent complex interdependences in utility functions, a number of representations are possible. In this section, we provide a generalisation of utility representation, that can be used to represent arbitrarily complex interdependencies in utilities of agents. Specifically, we will consider the negotiation occurring between two agents of *n* binary issues $x_1, x_2, \ldots, x_n$, where the negotiation determines the allocation of each issue, as well as a transfer (monetary amount) *p* that accompanies the deal. Intuitively, we can think of $x_1, x_2, \ldots, x_n$ as a choice over a bundle of items that will exchange hands, and to *p* as the price paid, but the representation is much more general, and the algorithm presented can be applied to any negotiation domain with binary issues.

**Linearly additive utilities:**

- Linear combination of issue utilities:

$$\mathrm{U}(x_1, x_2, \ldots, x_n) = \sum_{i=0}^{n} \alpha_i x_i \qquad \textit{(Eq. 1)}$$

Where *n* is the number of (binary) issues, $x_1, x_2, \ldots, x_n$ are the binary variables (allocations), and $\alpha_1, \alpha_2, \ldots, \alpha_n$ are the weights assigned to each issue

**K-additive utilities:**

- Represents a generalisation of additive preferences, that also captures local interactions (i.e. complementarity/substitutability effects) between groups of up to $k$ issues
- Have a polynomial representation:

$$\mathrm{U}(x_1, x_2, \ldots, x_n) = \sum_{i=0}^{n} \alpha_i x_i + \sum_{\substack{i,j=0 \\ i \neq j}}^{n} \alpha_{i,j}\, x_i\, x_j + \sum_{\substack{i,j,k=0 \\ i \neq j, j \neq k, i \neq k}}^{n} \alpha_{i,j,k}\, x_i\, x_j\, x_k + \cdots \quad \textit{(Eq. 2)}$$

  Where $n$ is the number of binary issues, $x_1, x_2, \ldots, x_n$ are the binary variables for each issue, and $\alpha_1, \alpha_2, \ldots, \alpha_n$ are the unary weights of each issue, $\alpha_{1,2}, \ldots, \alpha_{n-1,n}$ are the binary (2-additive) synergy weights, $\alpha_{1,2,3}, \ldots, \alpha_{n-2,n-1,n}$ are the 3-issue synergy weights etc.
- Here, we use $\boldsymbol{k}$ to represent the degree of the largest polynomial coefficient, for which not all $\alpha$ parameters are zero
- Such representations are much more difficult to deal with computationally. For example, even in a 2-additive case, it is much more difficult to find the optimal outcome than the linearly additive (or "1-additive") case for a large number of issues. For example, even for 100 binary issues, there may be up to 10,000 binary issue dependencies to consider.

A key observation to make is that the k-additive representation maps very well to a utility hyper-graph, where the k-rank dependencies are a weighted hyper-edge connecting k vertices. In this representation, the linearly additive case corresponds to a completely disconnected graph (no edges), where vertexes have a value corresponding to their weight, while the 2-additive case corresponds to a normal undirected graph (i.e. only 2-sided edges).

***Example 1: A simple utility graph***

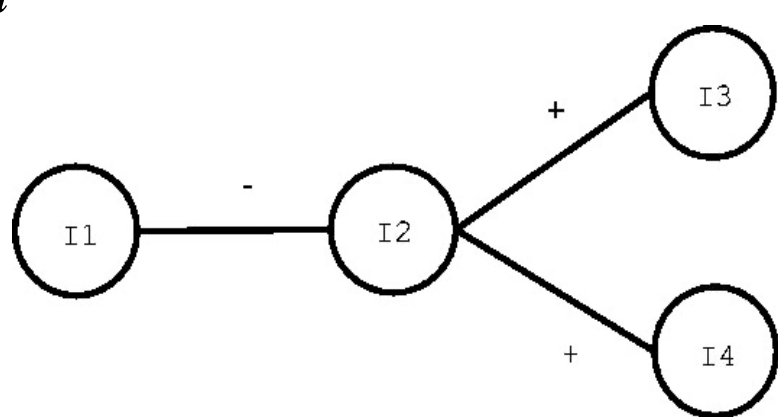

Consider $n=4$ binary issues and 2 agents, which intuitively, we call a Buyer and a Seller. The utility function of the Buyer is non-linear (as shown above): $U_B = 7x_1 + 5x_2 - 5x_1x_2 + 4x_2x_3 + 4x_2x_3$ and for the Seller it is linear, but negative: $U_S = -2(x_1 + x_2 + x_3 + x_4)$. In this case, the utility of the Seller is linear and negative (i.e. corresponds to a cost of 2 per issue), corresponding to an e-commerce scenario, but in the general case, the utilities of both agents can be complex and display interdependencies.
In this case, the Pareto-optimal bundles correspond to the contracts maximising the Gains from Trade (GT) from existing economics literature, defined as buyer's monetary utility gain minus seller's cost. For his example, the bundle maximising the GT is: $GT(I_1=0, I_2=1, I_3=1, I_4=1) = U_B + U_S = 13 - 6 = 7$.

### 2.1 Using utility graphs to decompose high-dimensional negotiation spaces

Given a negotiation problem where utilities have a k-additive representation, finding a Pareto-optimal outcome even for a small number of negotiating agents, is a complex problem. Consider a model in which both (or, possibly, more than 2) players have k-additive utilities $U_A$, $U_B$ and $U_C$, each of which represented by a k-additive utility function. In this case, a Pareto-optimal corresponds to an outcome – or set of outcomes that maximise $U_A + U_B + U_C$ (corresponding to a merged graph), noting that there can be multiple such outcomes. In fact, even finding the optimal outcome for one player can be difficult, if his/her utility graph is sufficiently dense, i.e. the number of utility constraint edges is high).

In the worst case, finding one Pareto-optimal outcome, in the binary issues case, requires for $n$ issues, $2^n$ steps (we can call these calls to the evaluation of the utility function, or "value queries" if we use the language of preference elicitation, discussed in Section 5). The key idea of graph decomposition is to identify a set of cut-set nodes (also called "interface nodes" by (Fujita et al.). These cut set nodes we need to consider every possible instantiation. But then we can search the optimal contract combination for each remaining subgraph or partition independently of the others, under the constraint that the values of the cut-set vertices are fixed. At the end, we find the overall optimal contract and cut-set combination, by comparing the sums of utility values of the local subgraph optima.

***Example 2: Achieving speed-up through graph decomposition***

Consider the following utility graph, assuming each node corresponds to a binary issue, and the edges correspond to a 2-additive polynomial constraint between that pair of issues:

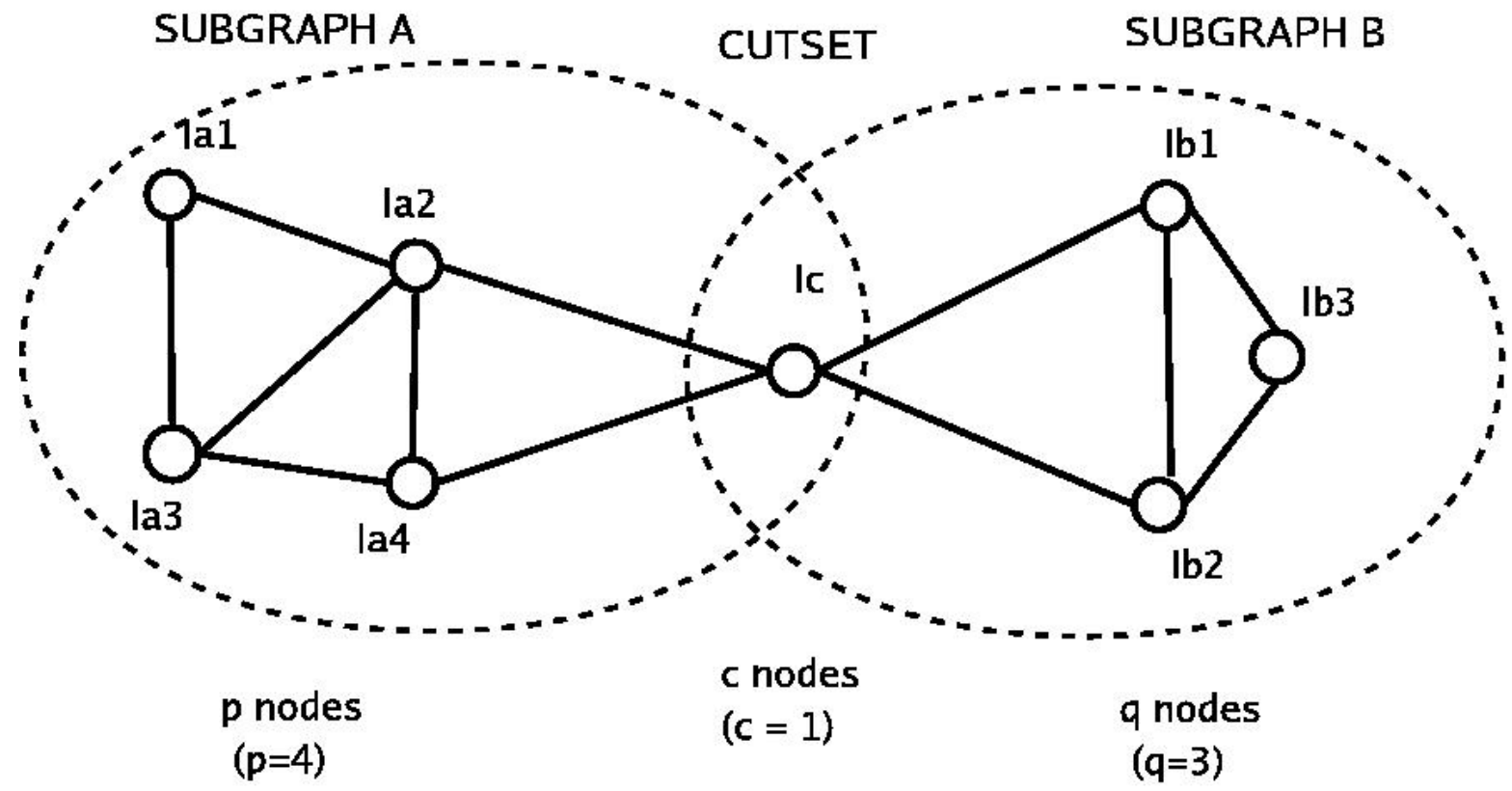

In this example (considering binary issues):

- If we do an exhaustive search for an optimal outcome, exploring all bundles: $2^{c+p+q} = 2^8 = 256$.
- If we use graph partitioning, we need to consider all the instantiations of cut-set issue (vertex) $I_C$, with either 0/1. Then, we can optimise separately to find the best combination in Subgraph A and Subgraph B for each possible value of $I_C$. At the last step we have to compute the optimal value of cut-set node $I_C$ buy comparing the value of the local optima in each of the 2 subgraphs, for each value of $I_C$. The complexity of this is: $2^c * (2^p+2^q) = 2 * (2^4+2^3) = 48$.

So, even for this simple 10-vertex graph we get a speed-up of 4 times. Note we could easily generalise the example, but e.g. having 10 nodes in both Subgraph A (p=10) and Subgraph B (q=10). In that case, the number of evaluations would reduce from $2^{21}$=2,097,152 to $2*(2^{10}+2^{10}) = 4096$ from having just one cut-set node (note that using more nodes or recursive procedure could decrease it significantly). Clearly, graph decomposition can result in exponential speed-up for very large graphs. However, a key challenge is, for a given utility graph structure at input, to **identify the decomposition** into convenient clusters of issues (subgraphs), such that the decomposition does not result in too many vertices assigned to cut-sets, which need to be explored exhaustively – yet at the same time the size of the resulting subgraphs after partition is considerably lower than that of the initial graph.

In fact, in terms of computation cost, the complexity of finding an optimal bundle is exponential in - and largely dominated by 2 factors: 1. The number of vertices in cut-sets, 2. The number of vertices in the largest subgraph (or cluster of issues) that remains after partition. This will be key into how we configure our graph partitioning algorithms.

## 2.2 Algorithms for search of optimal outcome using utility graphs

Consider a utility hyper-graph $G = \{V_1, V_2, .. V_n, E_1, .. E_m\}$ consisting of $n$ vertices and $m$ hyper-edges. This graph represents an interdependent utility function (see Eq. 2 above) as follows: each vertex corresponds to a binary negotiation issue, i.e. $V_i = I_i, \forall i = 1..n$ (where $I_i \in \{0,1\}$) and each (hyper)-edge $E_l, l = 1 \dots m$ corresponds to a weight $\alpha_{i,j,k\dots}$ between some issues (or vertices) $I_i, I_j$.. Note that $G$ is called a *utility graph* if there are no interdependencies of order higher than 2, otherwise it is a *utility hyper-graph*.
A ***decomposition*** (or *partition*) of the graph consists of a set of $k$ partitions or subgraphs $G_1 \cup G_2 \cup \dots G_k = G$, and a set of overlapping vertices $V_C$ that belongs to more than one subgraph, formally: $\forall i, j = 1 \dots k,\ (G_i \cup G_j) \in V_C$. The set $V_C$ is called the set of cut-set vertices as removing them would cut the initial graph into *k* separate, unconnected graphs.
Such a decomposition is called ***balanced*** (within some tolerance), if the difference in size between any 2 resulting subgraphs is bounded by some constant Ɛ, set to a low value (e.g. Ɛ=5% of the total number of vertices *n*). Formally we can write: $|G_i| - |G_j| \leq \varepsilon, \forall i, j = 1 \dots k$.
As the graph vertices represent binary issues, we can define an *instantiation* $\vec{x}_i \in \{0,1\}^n$ of graph $G$ an assignment of 0 or 1 to all issues (vertices) in $G$. For a binary graph, there are exactly $2^n$ such assignments. We denote an *optimal instantiation* as one of the assignments of values that maximises the total utility (note there may be several such assignments): $\vec{x}_i^* = argmax_{\vec{x}_i \in \{0,1\}^n}\, U(G(\vec{x}_i))$
Given this, we define the (very simple) brute-force search algorithm as:

**Algorithm 1 (Brute-force search):** $\forall \vec{x}_i \in \{0,1\}^n$ *return* $\vec{x}_i^* = argmax_{\vec{x}_i \in \{0,1\}^n}\, U(G(\vec{x}_i))$

The brute force search has a complexity of $2^n$, quickly becomes (e.g. $2^{50}$=$1.12 * 10^{15}$ and $2^{100}$=$1.27 * 10^{30}$, so processing larger graphs – as would be needed for even medium-complexity applications quickly becomes completely unfeasible.
Nevertheless, given a balanced decomposition, we can reduce the complexity considerably, by only having to explore exhaustively the values in the cut-set nodes. Then, for each possible assignment of cut-set node value instantiations, we can search for the optimal configuration in each subgraph independently (so all combinations of nodes for different subgraphs do not need to be considered).

**Algorithm 2: Optimal bundle configuration using graph decomposition**
*Given a partition of graph into k subgraphs* $G_1 \cup G_2 \cup \dots G_k = G$, *and a subset of cut-set of vertices* $V_C$ *representing all vertices that belong to more than one subgraph.*

*For* $\forall \vec{x}_C \in \{0,1\}^{|V_C|}$ *// for all binary combinations of vertices in* $V_C$

$$\vec{x}_i^* | \vec{x}_C = argmax_{\vec{x}_i \in \{0,1\}^{|G_i|}}\, U(G_i(\vec{x}_i)), \forall i = 1 \dots k$$

*//compute optimal assignment of vertices in each subgraph* $G_i$ *fixing the instantiation* $\vec{x}_C$ *of vertices in* $V_C$

$$U^*(\vec{x}_i^* | \vec{x}_C) = max_{\vec{x}_i \in \{0,1\}^{|G_i|}}\, U(G_i(\vec{x}_i))$$

*// max utility value of that subgraph for given* $\vec{x}_C$

*EndFor*

$$\vec{x}_C^* = argmax_{\vec{x}_C \in \{0,1\}^{|V_C|}}\, [U^*(\vec{x}_1^* | \vec{x}_C) + U^*(\vec{x}_2^* | \vec{x}_C) + \dots + U^*(\vec{x}_k^* | \vec{x}_C)]$$

*//select the cut-set combination* $\vec{x}_C^*$ *with the highest sum of optimal subgraphs utilities*

*Return:* $\vec{x}_C^*$ *and corresponding* $\vec{x}_1^*, \vec{x}_2^*, \dots, \vec{x}_k^* | \vec{x}_C^*$

It is not hard to see that the complexity of Algorithm 2, starting from a given partition $\mathcal{P}$ is:

$$O(\mathcal{P}) = 2^{|V_C|} \cdot (2^{|G_1|} + 2^{|G_2|} + \dots + 2^{|G_k|}) \quad \textit{(Eq. 3)}$$

It is useful to observe this can be upper-bounded by:

$$O(\mathcal{P}) \leq 2^{|V_C|} \cdot k \cdot 2^{Max|G_i|} \quad \textit{(Eq. 4)}$$

where the bound is exact for a perfectly balanced cut. This form shows more clearly that the computational complexity largely depends on the number of cut set nodes $|V_C|$, the total number of partitions $k$ and the size in nodes of the largest remaining subgraph $Max|G_i|$.
Let us consider how much more efficient the graph decomposition search in Algorithm 2 is compared to the brute-force search (Algorithm 1). For example, a graph with 100 binary issues can be divided into 5 subgraphs of 20 issues each with e.g. 6 cut set nodes, the complexity reduces from $2^{100}$ =1.27*$10^{30}$ to $2^6$ *5 * $2^{20}$ = 3.35*$10^8$. We can denote a simple speed-up metric of finding the optimal combination with the decomposed as:

$$SpeedUp = \frac{O(\mathcal{P})}{O(BruteForce)} \qquad \textit{(Eq. 5)}$$

So, for the above example, the speed up is 1.27*$10^{30}$/ 3.35*$10^8$ = 3.79*$10^{21}$
While the brute-force search is not useful in itself (and actually, it is simply not computable for most large graphs), it is a useful metric to measure the speed-up performance for different decompositions, depending on different types of graph topologies etc.

## 3. Graph-based decomposition algorithms: Implementation

In the prior section, we have proposed a new model and algorithms that allow us to computationally reduce the problem of finding Pareto-efficient outcomes in complex negotiation to one of graph partitioning. This is a key advantage, because the question of how to optimally partition a graph (or hypergraph) is a well-studied one in graph theory and optimization, as it appears in many practical settings, ranging from logistic networks, electronic circuit design, power network optimization etc. (see Newman [18,19] for an overview of this area).

In this work, we use *hMETIS* (Karypis and Kumar [13]), a free-source [hyper]-graph partitioning software package, that implements most existing state of the art graph decomposition techniques, and work for large-scale graphs. hMETIS has previously been applied to decompose graphs in complex real-life problems that involve large graphical representations, such as electronic circuit design, design of wireless networks, partition of road networks for traffic management, or partitioning datasets in large-scale data mining. Yet, to our knowledge, neither hMETIS nor any state of the art other (hyper)-graph partitioning solution has ever been applied before for complex, inter-dependent issue negotiation.

Use of hMETIS requires the user to specify the desired number of intended partitions, and aims to perform a balanced partition (i.e. where the partition sizes are relatively equal, up to a tolerable level of imbalance. In terms of representation, hMETIS work on both standard graphs (i.e. with 2-vertex edges), but also on hypergraphs, where an edge can link more than two vertices. Such hypergraphs with weighted hyper-edges correspond naturally to the k-additive representation of utility functions.

We found the key challenge we faced when using graph decomposition (and specifically hMETIS) for this problem is to specify the number of partitions we require. As already can be seen from *(Eq. 3)* above, there is a trade-off between having too few partitions, which would mean the largest one is still too large for exhaustive outcome computation, and too many partitions, which means the number of cut-set (interface nodes) becomes too large. We will examine this trade-off in the experimental analysis.

## 4. Types of utility graphs and their decomposability into issue clusters

### 4.1 Methods for generating complex utility graph structures

The complex systems literature studying theoretical and empirical graph structures (Newman'18 [18], Newman'03 [13], Strogatz [23], Barabasi & Albert [3]) identify three main types of graph structures that can emerge in complex physical, social and economic phenomena. These structures correspond to many real-life domains where automated negotiation could be applied, such as connections between people in social networks, distribution of loads in logistic networks, item buying from large merchants on the web (e.g. Amazon), power markets etc. The 3 main models for generating graphs that we considered in generating utility graphs in our negotiation work are:

- ***Random graphs:*** In random graphs, edges are relatively uniformly distributed in the graph; any vertex is as likely to be connected to any vertex as to the others (Erdös – Rényi model). There are several methods for generating them in practice. One is that an edge between any 2 nodes is chosen randomly with a (small) probability $p_i$. The size of this probability determines how dense the resulting graph will be. An alternative method of generating a random graph, given a fixed number of edges it must contain, is that the end point vertices of each edge are generated at uniformly random probability among the *1..n* vertices, avoiding any duplicates.
- ***Small-world graphs:*** Small world graphs have been observed in many real social networks and web networks/phenomena. They reflect the idea that vertices are more likely to be connected with others in their immediate vicinity, and increasingly less likely to be connected with others "further away" in the network. Small-world networks are generated by starting with a "locally connected" topology where each note is connected to only its neighbours and rewire each edge with a small probability *p* (Watts-Strogatz model). The intuition is that that most nodes know its immediate neighbours, but can meet new nodes with some random *p*.
- ***Scale-free graphs/networks:*** Scale-free graphs appear frequently in real-world systems, and have recently been observed empirically in a large number of applications (including web systems and e-commerce type applications). They are a very likely model for cases where utility graphs will be used in automated negotiation, such as electronic commerce and online advertising.
  Scale-free networks are generated through a preferential attachment (or "rich get richer") model (Barabasi–Albert model), where the most connected edges have higher probability of acquiring new connections. This result in highly concentrated networks, where some vertices are much more connected than others. A key property of scale-free networks is the power law distribution in edge connectivity, which leads to almost all vertices have at most 1-2 connections, but a very few vertices having a very large number of connections.

### 4.2 Graph decomposition and application in negotiation

In our work, we considered the three above methods for generating utility graph topologies, and examined the number of vertices required by hMETIS to separate them, as well as the result in number of evaluation calls required to find the optimal Pareto optimal contract.
In this section, to provide more intuition, we illustrate 3 random utility graphs of each type, showing the cut-set (or interface) vertices for each type. The **density** of each graph is the same: 100 vertices (corresponding to binary issues) and 100 edges (corresponding to binary, 2-issue dependencies). All the visualisations were done in Pajek (Batagliej et al. [4]), a free state of the art graph visualisation software.

Results are presented in Table 1 below. From these results, we can see that scale-free graphs (and to some extent small-world ones) are much more likely to be decomposable and hence useful for negotiation. We can see this intuitively in the graphs: In Fig. 5, if just 3 cut-set vertices labelled in green are removed (out of 100), very few of the largest subgraphs remaining would have more than ~5 vertices, while in Fig. 4, hMETIS finds 7 vertices to achieve a similar effect, whereas this requires considerably more vertices for uniform random graphs (Fig. 3). Moreover, some results for finding the Pareto-optimal outcome for the 50-issue, 50 edge case (after optimal separation) is given in the table below – each averaged over 10 graphs/instances.

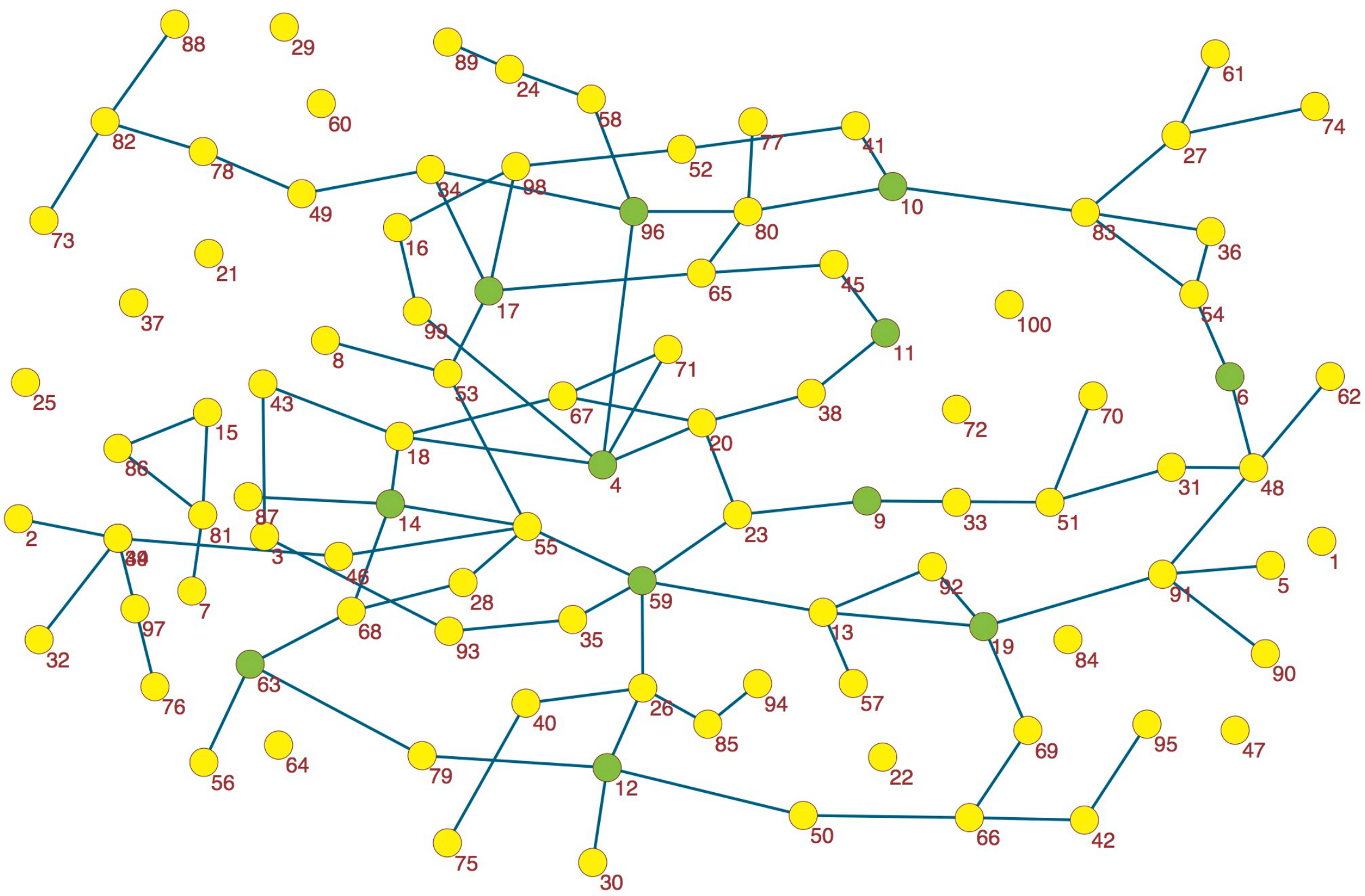

*Figure 3: Random graph with 100 issues and 100 binary dependencies (edges). Interface (cut-set) issues required to separate it into 10 components (subgraphs) are shown in green (12 issues/vertices out of 100)*

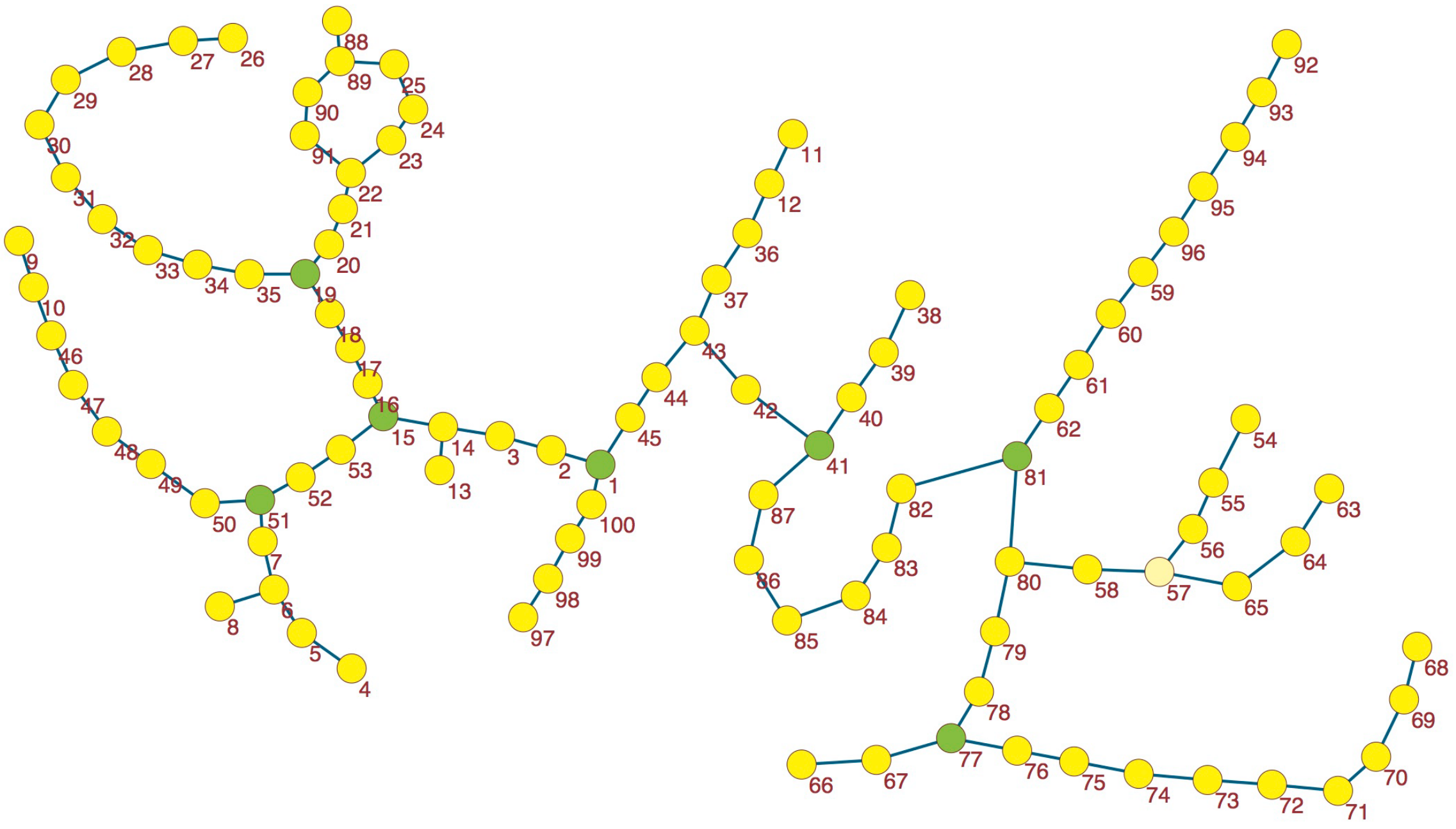

*Figure 4: Small world graph with 100 issues and 100 binary dependencies (edges). Interface (cut-set) issues required to separate it into 10 components (subgraphs) are shown in green (7 issues/vertices out of 100)*

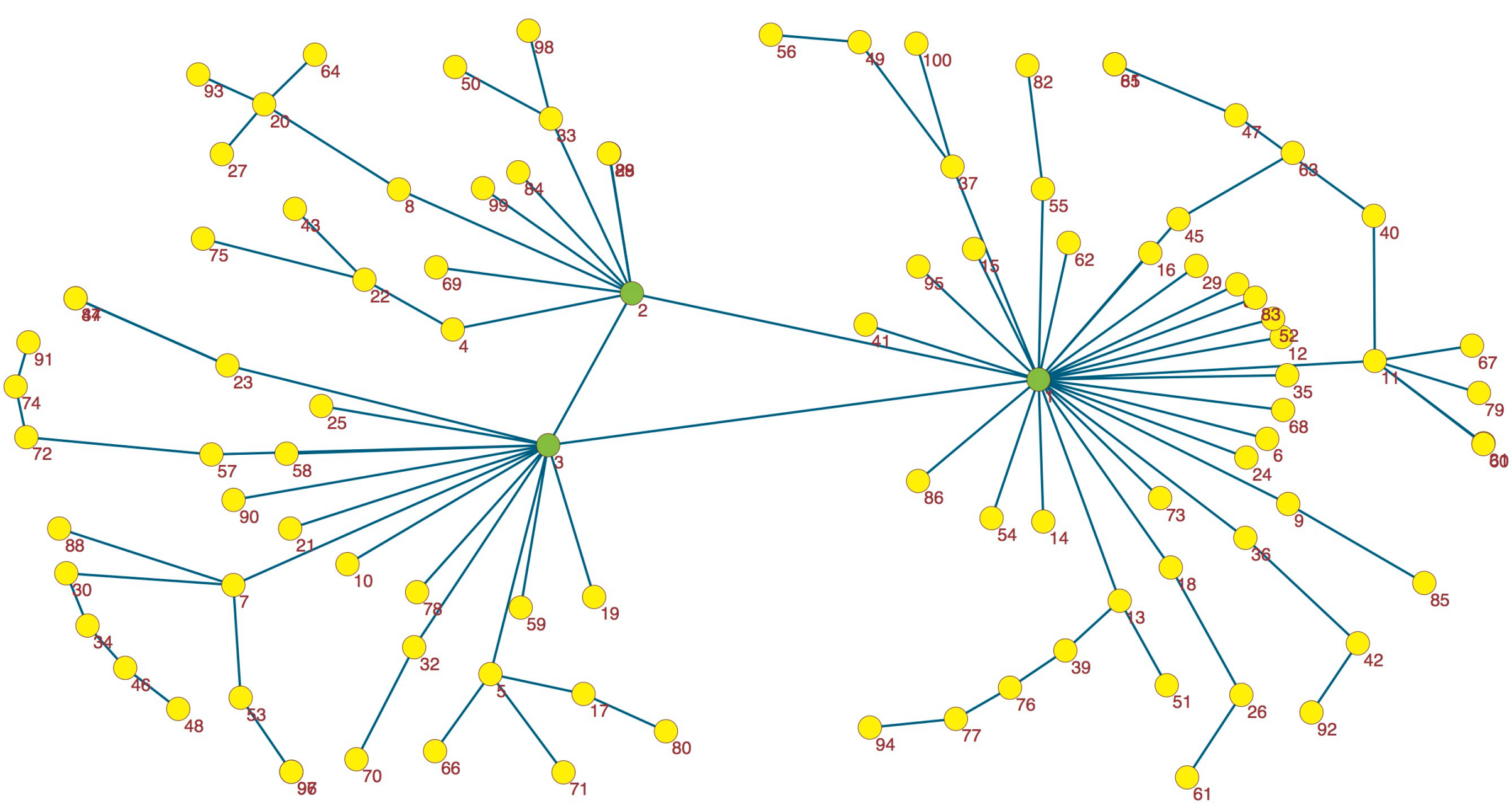

*Figure 5: Scale-free graph with 100 issues and 100 binary dependencies (edges). Interface (cut-set) issues required to separate it into 10 components (subgraphs) are shown in green (3 issues/vertices out of 100)*

| | No of utility function calls/queries to find Pareto outcomes (mean/std error.) | No. of interface vertices in the cut set (range) |
|---|---|---|
| Scale-free graph | 60,669 +/- 4,572 | 2-4 |
| Small world graph | 142,336 +/- 9,356 | 4-5 |
| Random graph | 806,912 +/- 35,16 | 8-10 |

*Table 1: Results for finding the Pareto-optimal outcome for a 50-issue binary graph, with the average vertex degree of 2 (i.e. 50 edges for 50 issues). For each case results are averages over 100 random graph configurations.*

From these experiments, we can highlight some key effects:

- It is possible to use graph partitioning algorithms to enable search for a Pareto-optimal solution in fairly large spaces with binary issues (with hundreds, potentially even thousands of binary issues). This result (i.e. applicability to large, random graphs) was not previously possible, given the prior state of the art. Both the prior work of Robu et al. [21] and Fujita et al. [11] only studied graphs with a maximum of 50 binary issues, and as far as we are aware, they applied their [mediated] negotiation algorithms only to specific graph structures chosen by hand, rather than generating a substantial number of random graph structures with different densities/configurations and examining whether they can be partitioned and used in negotiation.
- Another key observation that utility **graph density** (in terms of number of edges/constraints) is very important probably more important than size. For the above generated utility graphs, we used an average vertex degree = 2 (i.e. each vertex is, *on average*, connected to 2 edges). This means an equal number of vertices and edges in the graph (e.g. 100 vertices and 100 edges drawn at random, or 50 vertices and 50 edges). More dense graphs are possible, but become considerably harder to solve, while graphs that are close to fully connected are unlikely to be separable, and computing the Pareto-efficient outcomes would involve [close to] brute-force search. Nevertheless, real life utility graphs are highly unlikely to have as many constraints.
- Finally, the structure of the graph is a key criteria. As we can see from the above results, the search is exponentially harder in random graphs, compared to scale-free graphs (already for a 50-issues graph, by an order of over 10 times), while for larger graphs, this exponential gap only grows. In fact, a key observation is that, if preferences are structured close to a random graph, it becomes computationally expensive to find the Pareto-optimal outcome for more than 100 issues and dense graphs. Conversely, as we show in Section 6, for scale-free graph structures, we can handle interdependent issue negotiations of thousands or even tens of thousands of interdependent issues. In real life, however – in domains such as web services, e-commerce, transportation networks etc., scale-free graphs are much more frequently observed than graphs with uniformly-random distributed utilities, which makes our results promising to complex negotiations in many domains.

### 4.3 Optimal graph decomposition: cost of computation for different graph topologies

The above discussion already illustrates the importance of graph topology. In this section, we perform experiments to try to partition (in a balanced way) for larger utility graphs – of 1,000 binary issues (vertices), which is an order of magnitude larger than any utility graph considered in work so far - e.g. [21, 22, 11], but such a size is highly relevant in many domains where complex automated negotiations are applicable (e.g. logistic networks, city planning, online allocation systems etc.).

Specifically, here we look at the problem of computational complexity of the search based on the resulting balanced partition. In our theoretical analysis, we already highlighted a potential trade-off between having more, smaller partitions (hence the search of the optimal contract *inside* each partition is reduced), but at the rapidly increasing cost of having more separator (cut set issues/vertices).
Our analysis involved generating 100 graph configurations of 1,000 vertices each, for each of the 3 types of graph generation methods discussed: uniform random, small world and scale free graphs. Then, for each of these graphs, we compute the optimal balanced partitions, for all settings from 5 partitions (of average size of 1000/5=200 vertices/partition) to 150 partitions (average size 1000/150=~6 vertices/partition). Results are presented in Figure 6 – recall that each point in Figure 6A and 6B refers to an average over 100 different graph topologies).

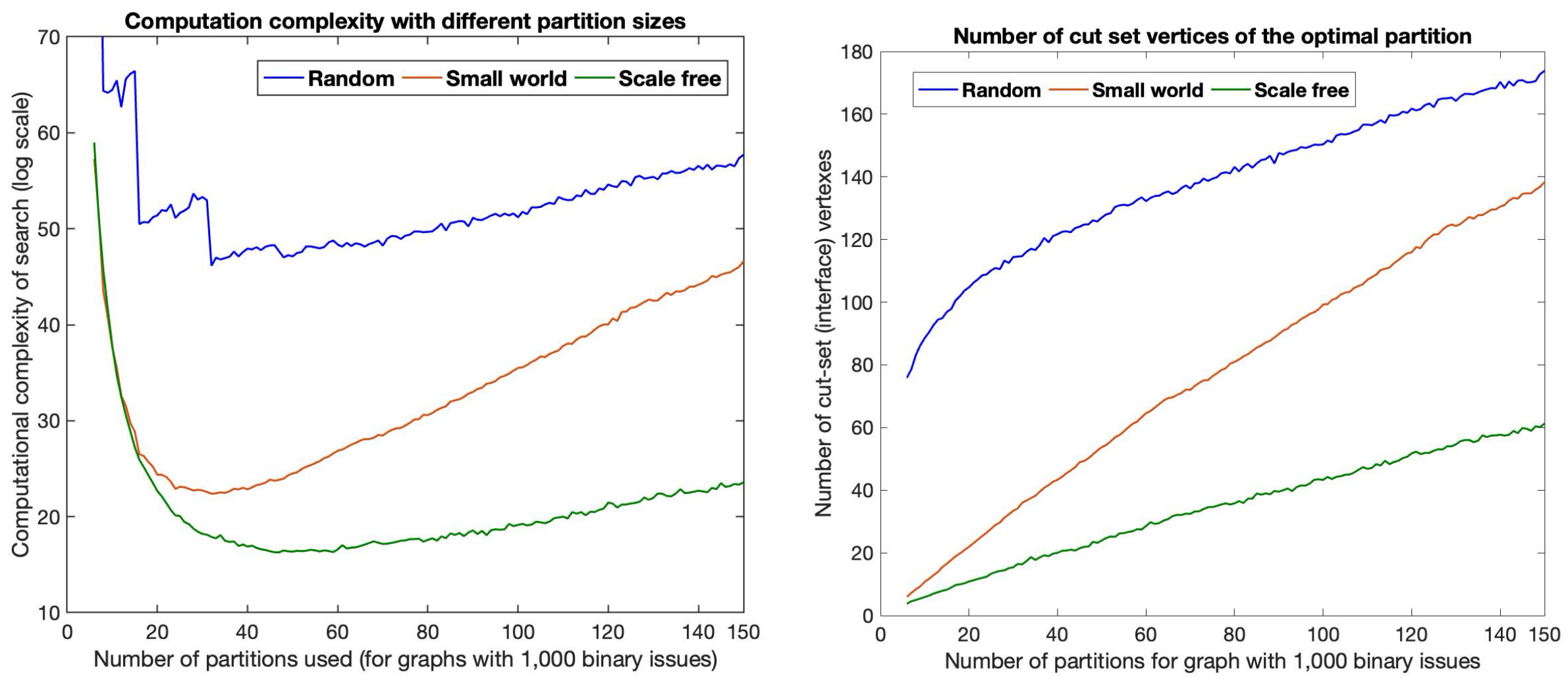

*Figure 6: Results from balanced partitioning utility graphs with 1,000 binary issues (vertices), into a number of subgraphs from 5 to 150 (hence the average size of the individual graph ranges from ~6 to 200). Left hand-side (A) shows the computational complexity of the global search for the optimum, while the right-hand side (B) shows no of cut-sets*

Examining these results, in Fig. 6A, which shows the total computational complexity of the search, computation is very expensive on the left side of the X-axis, where the number of subgraphs is small, but then it drops very fast. The reason is that for less than 20 partitions average size of the post-partition subgraphs remains too high, hence it is still expensive to compute the optimal combination in each subgraph. A minimum point is reached around ~40 partitions, but then starts to slowly increase, as we further increase the number of partitions. The reason for this can be seen in Fig. 6B – the number of cut-set interface nodes required to achieve partition also starts to increase fast, and the algorithm is exponential in them. So, there is a minimum at ~40 partitions (average number of vertices per partition of ~25), where the trade-off between number of partitions and number of cut set vertices needed to achieve that partition is optimal.
These results also make clear that graph topology has a big effect on the computational cost of search for Pareto-optimal outcome, even if utility graphs that have identical number of constraints (edges). While scale-free, and to a lesser extent small-world graphs are easier to partition even for very large sizes, completely random graphs are harder, because the number of cut-set nodes needed remains high. This is not unexpected, but on the positive side, we know many social phenomena that can lead to graphical utilities exhibit small-world and scale-free structures. In this work, we provide a general-purpose methodology for finding Pareto-optimal agreements, which can be applied to a variety of k-additive utility domains, but the analysis will need to be made on a case by case basis.

## 5. Mediated multi-issue negotiation protocols and link to preference elicitation

The main focus of the results presented so far has been on graphical utility representation and search for Pareto-optimal outcomes, without reference to a specific negotiation protocol. In terms of protocol, negotiation usually takes the form of bilateral bargaining, i.e. exchange of offers and counter-offers until an agreement is reached. Yet, for complex multi-issue negotiations, a mediator (i.e. independent party that performs the search/recommends possible outcomes) is almost always needed to find jointly profitable outcomes under complex multi-issue utility constraints. We also implemented our algorithms through a mediator - similar to the seminal work of Raiffa [20], but also most recent works on automated negotiation with complex interdependent issues (Ito et al. [12], Fujita et al. [11], Marsa Maestre [17]).

In more detail, in a mediated negotiation, a mediator will pose a number of queries/ask for bids iteratively from the agents, while running algorithms to search for jointly profitable outcomes. For example, such a query could ask each party what value they would assign to a specific instantiation of the binary items $I_1 ... I_n$ or ask them to compare two instantiations (see more about queries below). Mediated negotiation is still rather different than other "direct revelation" mechanisms, such as combinatorial auctions. In a combinatorial auction, the center asks all the parties for all their private utility information (utility function and weights), and computes the optimal allocation centrally, in a single step. By contrast, mediated negotiation occurs in iterations, and the mediator does not require each agent's private utility information in advance, but elicits their preferences in an iterated fashion, until a Pareto-optimal outcome is found. By contrast, we see mediated negotiations as closer to preference elicitation literature in AI, and we can use similar concepts, such as query types in this work.

### 5.1 Preference elicitation and query types

As discussed above, mediated negotiation has a close connection to preference elicitation, which has a long tradition in the AI literature, e.g. (Boutilier et al. [6], Brafman et al. [5], Lahaie & Parkes [16], Conitzer [10]), but is still rather different from it. In preference elicitation, the goal is to ask as many value/demand/comparison queries to determine the preferences of a single agent with high accuracy. By contrast, in mediated negotiation, the goal is not for the mediator to learn exhaustively the preferences of a single agent, but to find an outcome that is close to Pareto-efficiency for both (or all) parties. However, the framework of preference elicitation in AI provides us with some natural classes of queries to consider. In this work, we consider two classes of queries, that are well established in preference elicitation literature:

- **Value queries:** Such queries as the specific utility value associated with a specific contract/issue instantiation. E.g.: What is your utility value for ($I_1$=1, $I_2$=0, $I_3$=1)?

- **Comparison queries:** Such queries do not ask for specific utility value, but only for an order. Is your value for ($I_1$=1, $I_2$=0, $I_3$=1) greater than for ($I_1$=1, $I_2$=1, $I_3$=0)? (yes/no)

### 5.2 Use of queries by the mediator in negotiation

The key aim of the mediator, in our model, is to find the bundle that maximises the social welfare, i.e. the sum of the utilities of the negotiating agents. For a two-agent model (agents A and B), the goal of the mediator is to find a contract combination that maximises $U_A + U_B$ – however our mediated, utility graph-based negotiation model can be applied to more than 2 agents.

The mediator first partitions the joint utility graph of the constraints of both agents (consisting of the overlapping utility graphs), and identifies the set of cut set (interface) vertices using hMETIS, as described in Section 3 above. Denoting by $N$ is the set of all issues, and by $C \subseteq N$ the set of cut set nodes, and by $P_1, P_2 \ldots P_m \subseteq N$ the resulting partitions (subgraphs), such that $P_1 \cup P_2 \cup \ldots \cup P_m \cup C = N$ but partitions ***only*** overlap with each other through a cut-set node, i.e. $P_i \cap P_j \subseteq C, \forall i,j = 1..m$.

Next, for all vertices $\forall I_C \in C$ we consider all possible instantiations of $I_C \in \{0,1\}$. Then, iterating *for each set of instantiation* of vertices (issues) in C (there are $2^{|C|}$such instantiations):

- For **value queries,** for every partition $P_1, P_2 \ldots P_m$ we compute the instantiation that maximises $U_A + U_B$, *conditional on* that nodes in each partition that belong to the cut set ($\forall I \in C \cap P_i$), their values are fixed
- Then, we compute the global optimum by determining for which combination of nodes $\forall I_C \in C$, the sum of the local optimal utilities found for each partition (subgraph) $P_1, P_2 \ldots P_m$ is maximised.

The complexity of this procedure (for binary-values issues) is: $O(2^{|C|}(2^{|P_1|} + 2^{|P_2|} + \cdots + 2^{|P_m|}))$ as opposed to $O(2^{|C|+|P_1|+|P_2|+\cdots+|P_m|})$ for a brute search approach. Note that for the value queries case, we can compute the value of the optimum in each subgraph, and then determine the configuration of cut set nodes for which the sum of subgraph maxima is maximised.

**Comparison queries**

For the comparison queries case, because the mediator only receives order information, the mediator algorithm can compute the set of contracts on the Pareto-efficient frontier through comparison queries, even if it does not know the actual utility value (which remains private to the agents). Algorithm steps:

| U(Agent_0) ⇩ | Util(Agent_1) ⇩ |
|---|---|
| **$C_3$** | **$C_6$** |
| **$C_4$** | $C_1$ |
| $C_5$ | **$C_4$** |
| **$C_6$** | $C_2$ |
| $C_2$ | $C_5$ |
| $C_1$ | $C_7$ |
| $C_7$ | **$C_3$** |
| $C_8$ | $C_8$ |

**Step 1**: Partition the graph using hMETIS, as before, and consider all $2^{|C|}$ instantiations with $\{0,1\}$ of cut set vertices in $C$

**Step 2:** For each subgraph $P_i$ we can generate $2^{|P_i|}$ possible contracts but keeping only those respecting the instantiation of $C$. Then the set of Pareto-optimal contract can be generated as (see figure):

- Order contract descendingly according to both agents' utility
- Add contracts to the Pareto set descendingly, discarding any

contracts already dominated by a preferable contract for the other agent (e.g. $C_5$ is dominated by $C_4$, any contract below $C_6$ must be dominated)
This has a complexity ($N_C$ is the no of contracts in the partition) $O(N_c(1 + 2 \log N_C))$ (requires two sorts + one transversal)

**Step 3:** Merge, by comparison queries of local Pareto optimal contracts for all cut set node combinations $C$

The complexity of the procedure depends on local no of Pareto points/partition and not cut set nodes, but it is at least one order of complexity higher than for value queries.

## 6. Results for computational speed-up achieved by our algorithm

In this section, we discuss some of the experimental results for a number of increasingly larger randomly-generated graphs (both in terms of structure and edge weights). For these results (so far), we looked at 2-additive graphs (only 2-issue edges), where the density is given by an average vertex degree of 2 (which means graphs where the number edges equals the number of vertices/issues). The results shown here are for randomly-generated scale-free graph structures.

Since we are always interested in the complexity of reaching Pareto-efficient agreements, the main criteria we use to evaluate the performance of our experiments is the speed up obtained from graph decomposition and search algorithms we develop. The benchmark used for speed-up is always measure w.r.t. the exhaustive search, i.e. querying the value of the $2^N$ bundles for *N* binary issues. Also, to keep the computation cost independent of the machine on which the code is run and programming language, speed is measured in the number of *queries* that the mediator makes to the participating agents. Specifically, these are the number of calls to the utility evaluation function for value queries case, or the result of a comparison between the utility of two contracts for comparison queries case. Formally, speed-up is defined as:

$$Speedup = \frac{Number\ queries\ with\ graph\ decomposition}{Number\ of\ queries\ exhastive\ search} \quad \textit{(Eq. 6)}$$

Figure 6A shows speed-up results for randomly generated scale-free graphs and utility values, for sizes ranging from 20, 50, 75 and 100 binary issues. For all these negotiations, the graph decomposition method was run until end of the mediated negotiation (i.e. until the set of Pareto-optimal solutions was found). Note that the Y-scale is ***logarithmic*** (i.e. shows a $\log_2()$ of the actual speed up result, e.g. $2^{58}$ $=10^{17}$ speed-up). Obviously, for anything beyond 20 binary issues, the exhaustive queries search method is completely unfeasible to run in practice, although we know the number of queries is $2^N$). Note that the difference in speed-up from value queries to comparison queries case looks small and constant, but this is due to the exponential scale of the Y-axis. In fact the difference is not constant and exponential, ranging from $2^2$-$2^{2.5}$, or 4X to 5.5X times more queries needed in the comparison queries case.

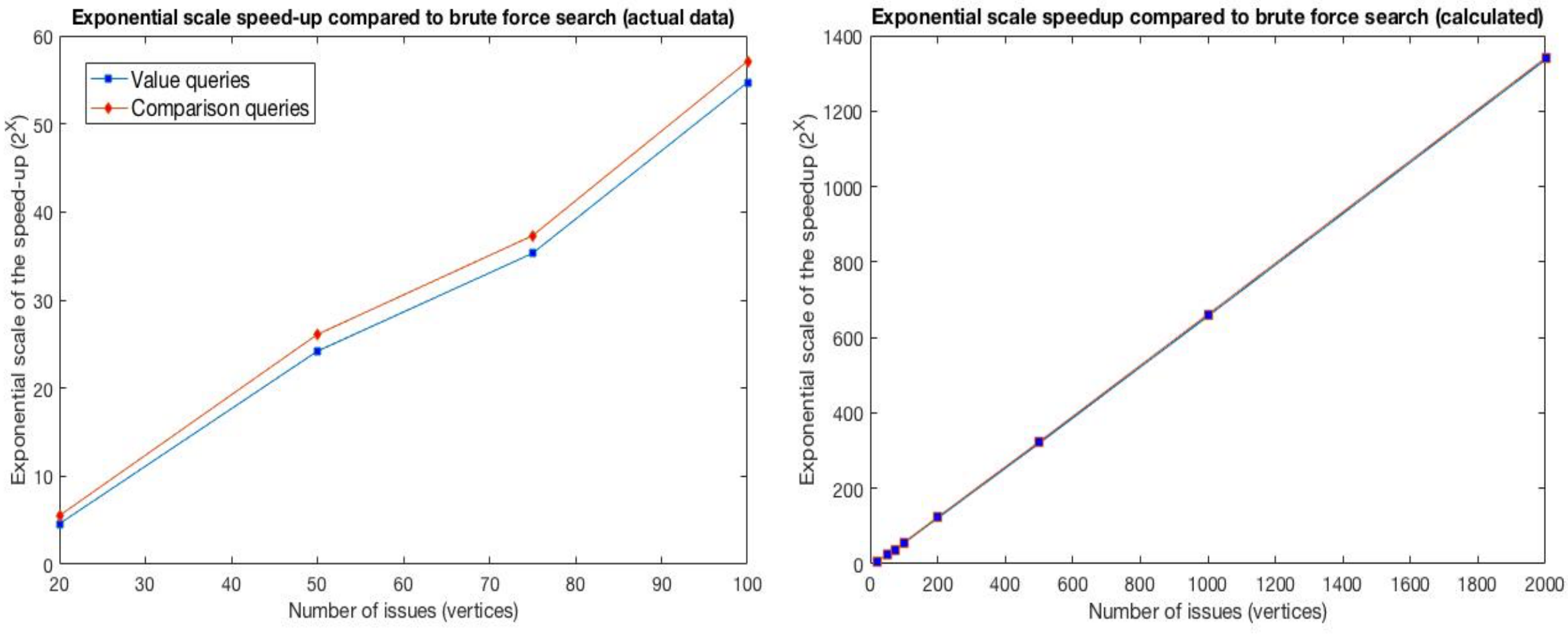

*Fig 6A (left): Comparison of speed-up achieved in negotiations using value vs. demand queries for scale-free utility graphs of up to 100 vertices. Fig 6B: Calculated speed-up in negotiations with random scale-free graphs of up to 2000 vertices*

Next, we wanted to look at even larger graphs, up to 2,000 binary issues (Figure 6B). These large configurations (essentially any configuration beyond 200 binary vertices) take over 15-30 minutes to run on a standard MacbookAir laptop, but they would be much faster on more powerful machines. Results in Figure 6B focuses on scale-free graphs and value queries case, for which we can compute the computational complexity (in terms of number of queries needed) with high accuracy, even if we don't run the full negotiations until the Pareto-optimal configuration is found. Hence, in Figure 6B, the speed-up results for 20-200 issues are generated from fully concluded negotiations, for 500, 1000 and 2000 binary issues it is estimated based on the computational complexity of the value queries procedure. This graph clearly shows the power of graph partitioning in negotiation: for larger graph sizes, it become completely unfeasible to compute Pareto-optimal outcomes, except through graph partitioning.

Finally, in Figure 6C, we look at some of the underlying reasons for the exponential speed-up in search. Specifically, we look at how the number of cut-set (interface) vertices after decomposition scales up as the number of vertices in the graph increases (recall that it is the number of cut-set vertices and size of large resulting subgraphs that drives search complexity). Given the capabilities of hMETIS, we can do this for randomly-generated scale-free graphs up to 10,000 issues/vertices – with an average vertex degree of 2 (no of generated edges equal to number of vertices). Here, we set the target size of the subgraphs resulting from the partition to 20 (+/-5) vertices. From Fig. 6C we can already see that, even for large graphs, the number of cut-set vertices is still only 2.5% of the total, which points to the fact that even these large graphs are decomposable, hence negotiations on them potentially tractable (note however, this result applies to scale-free graphs, for random graphs, the maximum tractable size would be much more restricted).

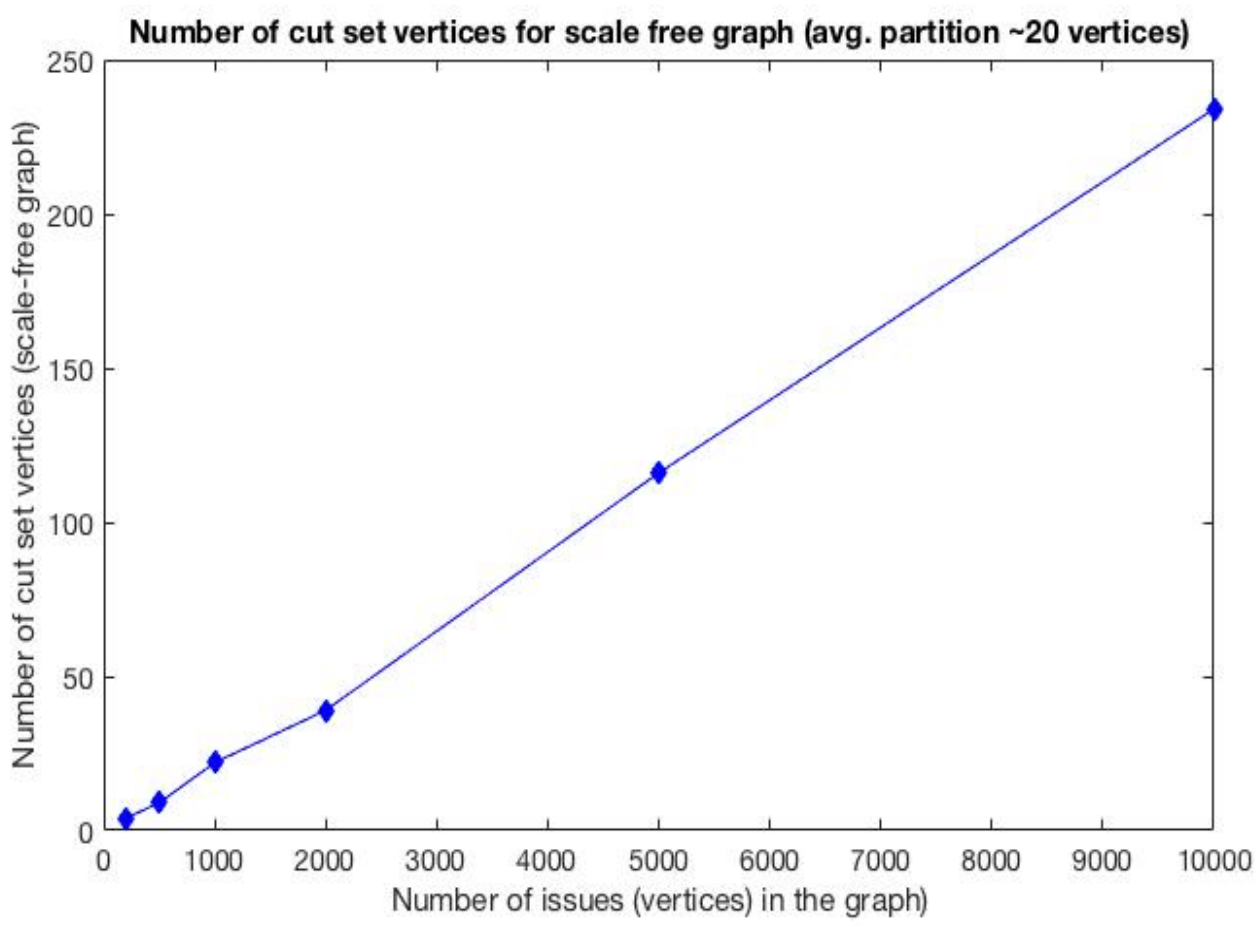

*Fig 6C: Number of vertices in the cut-set ("interface" vertices) in scale-free graphs with up to 10,000 issues*

## 7. Summary of Contributions & Further work

This work makes several key contributions to existing state of the art in complex automated negotiation with interdependent issues. First, we provide a new graphical formalism for representing utility functions with interdependent, k-additive utilities, as well as a number of algorithms to search the solution space for outcomes maximising the Gains of Trade (hence, Pareto-efficient). This approach allows us to algorithmically reduce the problem of computing Pareto-efficient outcomes in mediated negotiation to one of graph partition - a problem which is well studied in graph theory.

The proposed solution allows us to model negotiations with much larger utility spaces (i.e. many more issues and dependencies) than was previously possible in the work of Klein et el. [14], Fujita et al. [11], Robu et. al. [21]. Also, by contrast to other works, our method is guaranteed to converge on finding Pareto-optimal outcomes, unlike other works that find heuristically good outcomes, but do not guarantee an optimal solution. We also provide a principled method to examine the trade-off in a balanced partition of a utility graph used in negotiation between having too few partitions (leaving the large partition too large) and too many partitions (hence too many cut-set vertices), with an optimal trade-off point for each type of graph, that minimises the computation costs.
Second, our work is the first to discuss a number of ways to generate utility graph topologies for negotiation modelling, following established models to generate random graphs from complexity science. We show that the method of generating a random utility graph topologies plays a crucial role (even more that raw size) in how decomposable the graph is, and hence how difficult it is to find Pareto-optimal contracts. Some structures (e.g. scale-free graphs) lend themselves much better to decomposition and use for search of Pareto-optimal outcomes in negotiations than others (e.g. graphs with uniformly distributed edge connectivity). For some structures and graph densities, we show our algorithms can handle negotiations over very large utility graphs, while for others (uniform random graphs) we can still handle relatively large negotiations, although the problem becomes harder to solve quicker, as the utility graphs are less decomposable. This provides important insights for researchers that aim to apply automated negotiation with interdependent utilities solutions in a particular domain (e.g. electronic commerce, logistics, transportation networks, ridesharing etc.), because it shows that a study of the underlying graphical preference structure can leveraged to design efficient negotiation algorithms. In fact, given recent observations in complex systems science, scale-free graphs are likely to appear in a wide variety of practical domains of interest in negotiation, such as emerging web applications (crowdsourcing, e-commerce), making them easier to decompose.

Finally, our work also makes a link between mediated negotiation with interdependent issues and the work on preference modeling in AI. We show how finding Pareto-optimal outcomes for many classes of utility graphs can be achieved, using a number of query types (specifically value and comparison queries) from the preference elicitation in AI literature. Establishing this link between the areas of mediated complex negotiation and preference elicitation query types potentially makes our results of broader interest in the AI preference modeling community.

In terms of future work, several directions would provide promising extensions.
First, so far, our work only deals with binary issues - we argue this is sufficient to illustrate the power of the method, following several other well-established papers on this topic. It is fairly straightforward to extend the model and algorithms presented to a model in which each issue has a number of discrete levels (as opposed to just two, as the binary case). Yet, for more complex cases, such as issues being represented by intervals or continuous variables, the model would require more complex extensions, and making connections to techniques from other areas such as computational geometry (Klein [15]).

Second, so far, the tests we performed considered two-agent negotiation. We believe the model can be extended to search for agreements with more than 2 agents, where the goal is to search the agreement maximising the sum of the agents' utility functions. Having more agents would clearly have a cost in terms of the search and number of constraints, but the speed-up achieved would not be too different. But, for really complex cases of negotiations between many agents (e.g. ridesharing, vehicle routing in logistics etc.), we envisage that the agents themselves (not just the issues under negotiation) could be grouped into negotiating sets or coalitions first, based on some underlying graphical relationship model.

Finally, looking further ahead, we are aware that utility graphs and k-additive utilities are not the only way to represent interdependence relations in complex, interdependent issue negotiations - and many other types of constraints exist, especially for real-valued issues (e.g. integral constraints, geometric constraints, cone constraints etc. (see (Marsa-Maestre et al. [17], Fujita et al. [11])), and also many other computational methods to address them, such as those from computational geometry, constraint satisfaction, evolutionary algorithms etc. It would be interesting to see how these could be compared to, and integrated with graph-based approaches. We argue the international Automated Negotiating Agents Competitions is a good venue to bring together interests in this community (see Aydogan et al. [1,2] for a recent overview of ANAC). Although currently, most of the negotiation competition tracks use linearly additive functions, some efforts have already started on the interdependent issue case, and we hope our work will contribute to future developments of this track on ANAC.

Overall, we conclude that through the algorithms and test methodologies proposed in this paper, we have shown the power of utility graph representation has the ability to handle truly large and complex negotiation spaces, advancing the state of the art for this challenging computational problem.